\documentclass[sigconf, nonacm]{acmart}

\usepackage{multirow}

\AtBeginDocument{%
  }

\begin{document}

\title{Versioned Late Materialization for Ultra-Long Sequence Training in Recommendation Systems at Scale}

\author{
Liang Guo, 
Ge Song,
Litao Deng,
Jianhui Sun,
Chufeng Hu,
Lu Zhang,
Zhen Ma,
Shouwei Chen,
Weiran Liu,
Sarang Masti Sreeshylan,
Xiaoxuan Meng,
Yanzun Huang\textsuperscript{$\dagger$}
} 
\affiliation{Meta Platforms, Inc.
\country{USA}}

\renewcommand{\shortauthors}{Guo et al.}

\begin{abstract}
Modern Deep Learning Recommendation Models (DLRMs) follow scaling laws with sequence length, driving the frontier toward ultra-long User Interaction History (UIH). However, the industry-standard ``Fat Row'' paradigm, which pre-materializes these sequences into every training example, creates a storage and I/O wall where data infrastructure usage exceeds GPU training capacity due to data redundancy that is amplified in multi-tenant environments where models with vastly different sequence length requirements share a union dataset. We present a \emph{versioned late materialization} paradigm that eliminates this redundancy by storing UIH once in a normalized, immutable tier and reconstructing sequences just-in-time during training via lightweight versioned pointers. The system ensures Online-to-Offline (O2O) consistency through a bifurcated protocol that prevents future leakage across both streaming and batch training, while a read-optimized immutable storage layer provides multi-dimensional projection pushdown for heterogeneous model tenants. Disaggregated data preprocessing with pipelined I/O prefetching and data-affinity optimizations masks the latency of training-time sequence reconstruction, keeping training throughput compute-bound by GPUs. Deployed on production DLRMs, the system reduces training data infrastructure resource usage while enabling aggressive sequence length scaling that delivers significant model quality gains, serving as the foundational data infrastructure for modern recommendation model architectures, including HSTU~\cite{zhai2024actions} and ULTRA-HSTU~\cite{ding26ultra}.
\end{abstract}

\ccsdesc[500]{Information systems~Recommender systems}
\keywords{Recommendation Systems, Machine Learning, Late Materialization, Training Data Infrastructure, Sequence Modeling}

\maketitle

\def\thefootnote{\fnsymbol{footnote}}
\footnotetext[2]{Work done at Meta Platforms, Inc.} 
\def\thefootnote{\arabic{footnote}}

\section{Introduction}
\label{sec:intro}

The advances of DLRMs over the past decade have been fundamentally driven by the length of the User Interaction History (UIH). Early candidate-dependent search approaches (e.g., DIN~\cite{zhou2018deep}, SIM~\cite{pi2020search}, ETA~\cite{chen2021end}) select the top-K relevant history items per candidate and scale UIH from $10^1$ to $10^4$ events. More recently, HSTU~\cite{zhai2024actions} reformulates recommendation as a sequential transduction task, applying full causal self-attention over the entire interaction sequence to avoid information loss and pave the way for the Generative Recommender framework. Notably, the work established that recommendation model quality follows a power-law scaling relationship with training compute---providing the first evidence that the scaling laws driving the LLM revolution extend to recommendation systems. This finding has been corroborated across the industry~\cite{li2025fm, ding26ultra, chen2026massivememorizationhundredstrillions, si2024twin, lyu2025dv365, chai2025longer, liu2025longterm, ren2025longretriever}, with the consistent result that longer sequences yield monotonically improved recommendation quality---pushing the frontier towards $10^{5}$+ events.

However, these architectural advances have exposed a fundamental infrastructure bottleneck. The industry-standard approach pre-materializes complete UIH sequences into every training example, a ``Fat Row,'' causing $K$-fold data redundancy as the same user history is duplicated across all ranking requests within the lookback window. The resulting storage and I/O amplification drives the total resource usage of data supporting services to exceed the usage of training GPUs themselves, effectively capping the scaling potential of next-generation architectures.

We observe that this problem is based on a false necessity. The industry adopted physical pre-materialization as a blanket mechanism for Online-to-Offline (O2O) consistency---ensuring training reproduces the exact feature state observed at inference time. While O2O consistency is required, pre-materialization is only sufficient, not necessary, to achieve it. UIH sequence data possesses a structural property---it is an append-only, temporally ordered, immutable sequence---that admits a more efficient alternative: storing the canonical history once and reconstructing the exact inference-time state via a temporal predicate at training time. Although late materialization and multi-version concurrency control (MVCC) are well-established techniques in the database community~\cite{abadi2007materialization, bernstein1983mvcc}, their application to recommendation training data pipelines---where the interplay of streaming and batch training modes, future-leakage prevention, and multi-tenant sequence projection creates domain-specific challenges absent in traditional query processing---has not been explored. In this paper, we present a \emph{versioned late materialization} paradigm that exploits this insight and breaks through the storage and I/O wall. Our contributions include:

\begin{enumerate}
  \item A formalization of the ``Fat Row'' Problem, with quantitative analysis showing that pre-materialized UIH redundancy drives data infrastructure usage to exceed GPU training deployment in recommendation systems (Section~\ref{sec:motivation}).
  \item A \emph{versioned late materialization protocol} that achieves O2O consistency and prevents future leakage through lightweight version metadata, serving as a unified solution for both streaming training and batch training (Section~\ref{sec:methodology}).
  \item A \emph{production-grade training-time materialization architecture} combining read-optimized immutable storage with multi-tenant sequence projection pushdown, disaggregated data preprocessing, pipelined I/O prefetching, and data-affinity sharding to mask sequence reconstruction latency and keep training throughput GPU-bound (Section~\ref{sec:productionization}).
  \item \emph{Deployment at scale}, reducing training data infrastructure resource usage while enabling model quality gains through aggressive sequence length scaling, serving as the foundational infrastructure for HSTU~\cite{zhai2024actions} and trillion-parameter sequential transducers (Section~\ref{sec:evaluation}).
\end{enumerate}

\section{The Fat Row Problem}
\label{sec:motivation}
This section examines the fundamental infrastructure bottleneck that emerges when scaling UIH sequences under the industry-standard pre-materialization paradigm. We first describe the Online-to-Offline consistency requirement that motivates the ``Fat Row'' architecture, then quantify the storage and I/O wall, and finally show how multi-tenant training environments amplify the penalty.

\subsection{Online-to-Offline Consistency}

A defining challenge of industrial recommendation systems is that model training typically occurs asynchronously: seconds, minutes, or hours after the original ranking request, yet training must reproduce the exact feature state observed during online inference, to avoid a well-documented source of model quality degradation~\cite{sculley2015hidden} from O2O data skew. The most pernicious form of skew is \emph{future leakage}. In a typical lifecycle, a ranking request is triggered to retrieve videos for a user's next session at $T_{\text{request}}$. The user's engagement (the label) with one of the videos in the next session occurs at $T_{\text{label}}$ ($T_{\text{label}} > T_{\text{request}}$). When training on this session asynchronously at $T_{\text{train}}$, if a training example inadvertently includes user events that occurred after $T_{\text{request}}$, the model learns to exploit information unavailable at inference time, producing artificially inflated offline metrics that do not transfer online.

\begin{figure}[t]
  \centering
  \includegraphics[width=\linewidth]{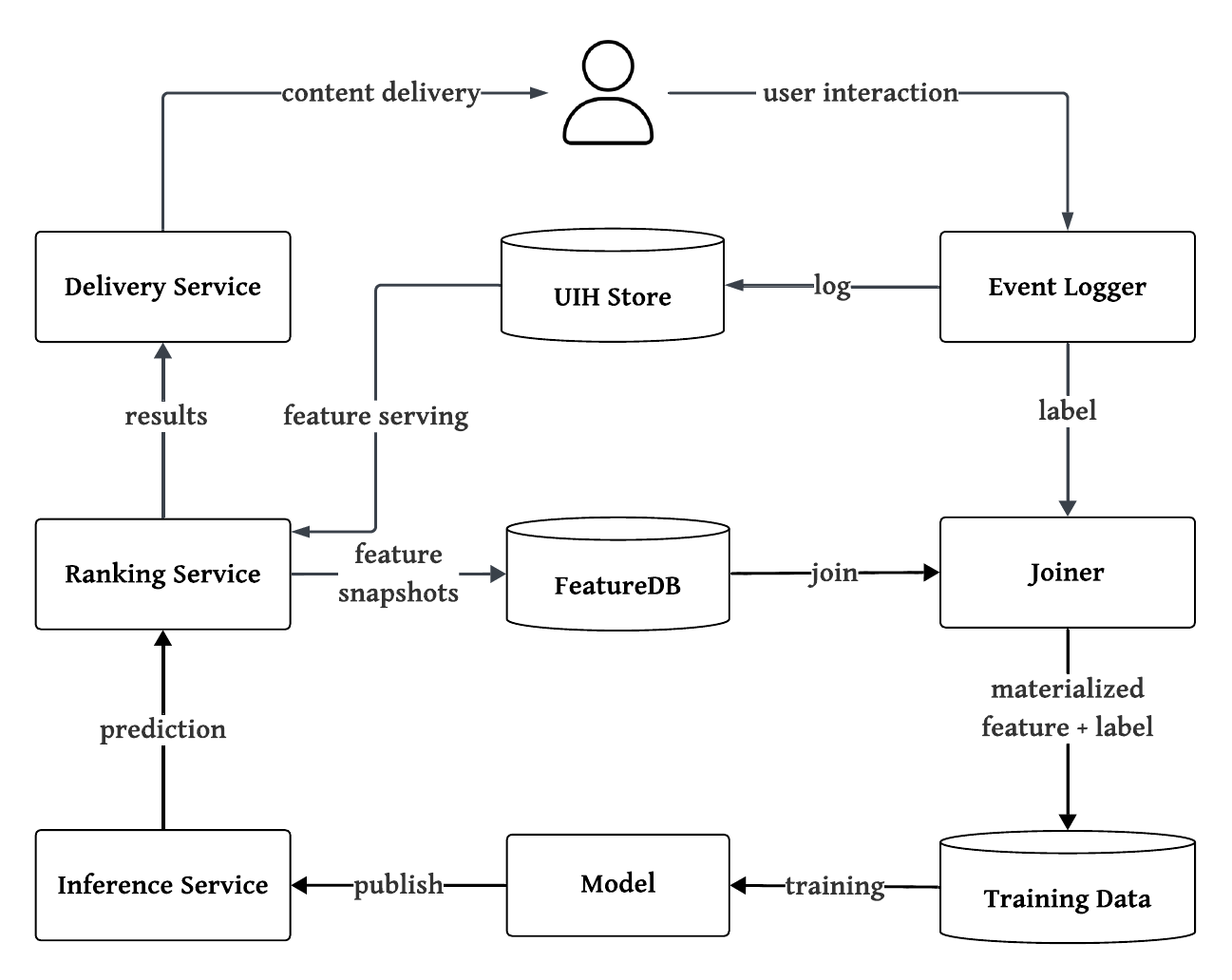}
  \caption{Feature snapshotting and pre-materialization architecture for RecSys training data generation.}
  \Description{Architecture diagram showing the feature snapshotting and pre-materialization pipeline.}
  \label{fig:feature_snapshot_architecture}
\end{figure}

To ensure that a model is trained on the exact state of features seen during online inference, the industry-standard architecture employs a feature snapshotting and pre-materialization mechanism illustrated in Figure~\ref{fig:feature_snapshot_architecture}. During the online inference funnel, the high-dimensional feature vectors used for ranking, including long-sequence UIH features extracted from the frequently updated UIH store, are ``frozen'' and cached in a high-throughput key-value feature store. As ground-truth labels from user interactions (e.g., likes or video completions) arrive asynchronously, an ingestion service joins them with the cached snapshots to generate a fully materialized training example~\cite{guo2025request} containing all feature values and associated late-arrival labels---a ``Fat Row.'' This architecture must serve both \emph{streaming} online training (where samples are consumed within seconds) and \emph{batch} offline training (where samples may be replayed days later), making the consistency guarantee non-negotiable. Although this architecture ensures high-fidelity data, it forces the system to store and transport redundant UIH sequence features for every individual training example, creating the infrastructure bottlenecks that our system aims to solve.

\subsection{Storage and I/O Wall}

As DLRMs evolve towards ultra-long sequences to capture lifelong user interests, the systemic redundancy of the ``Fat Row'' paradigm has become a bottleneck to model scaling. In this paradigm, training data are generated by physically denormalizing and materializing UIH features into every individual training example. The impact of this approach is best illustrated by the relationship between request frequency ($K$) and historical lookback ($N$). To provide the model with an $N$-day UIH window, the user's history for the preceding $N-1$ days---which remains static across $K$ requests in the same day---is physically duplicated into every generated training example. This results in a $K$-fold amplification in both storage footprint and write volume for the vast majority of the UIH payload.

\begin{figure}[t]
  \centering
  \includegraphics[width=\linewidth]{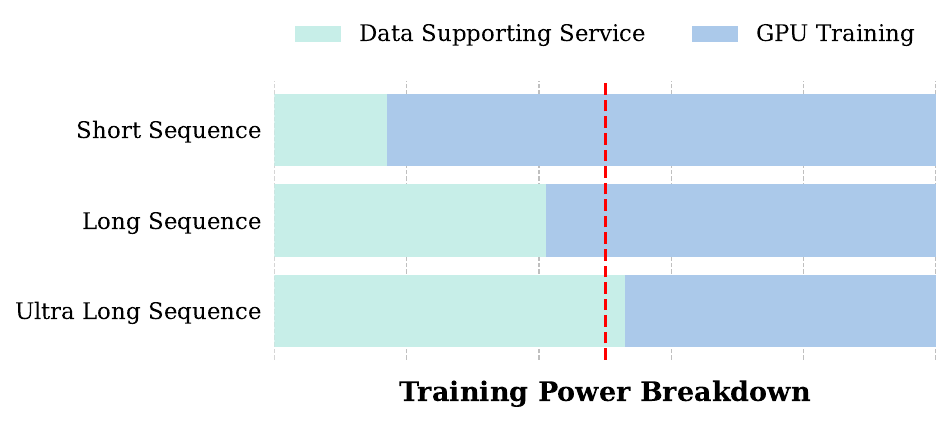}
  \caption{Estimation of data supporting service and GPU training power of a recommendation product to support short, long, and ultra-long sequences.}
  \Description{Bar chart showing the breakdown between data supporting services and GPU training.}
  \label{fig:power_breakdown}
\end{figure}

As shown in Figure~\ref{fig:power_breakdown}, our analysis of a recommendation surface reveals the severity of this wall. The total resource consumption of data supporting services---comprising training data storage, online and offline training data ingestion, and preprocessing compute---exceeds that of GPU training when enabling ultra-long UIH sequences. In this imbalance, UIH sequence features account for the majority of the training data volume and read bandwidth. The industry is forced into a ``diminishing returns'' trap where the infrastructure demands of supporting longer sequences eventually outweigh the marginal gains in model quality, effectively capping the potential of next-generation architectures.

\subsection{The Multi-Tenant Penalty}
\label{sec:union}
The Fat Row penalty is amplified in multi-tenant environments. In mature recommendation ecosystems, it is standard practice to maintain a union training dataset, a centralized dataset serving multiple model tenants across the recommendation funnel (e.g., Retrieval, Pre-ranking, and Ranking) of the same product. This unified approach offers significant operational benefits by minimizing storage redundancy and pipeline maintenance, and providing a consistent data baseline for cross-model performance ablation.

However, under the ``Fat Row'' paradigm, this shared infrastructure creates a multi-tenant penalty. Different models have vastly different requirements: a candidate retrieval model may only require the last 100 user events; a late-stage ranking model may require $100\times$ more events to maximize precision. Because the union dataset must accommodate the maximum requirement, it materializes and writes the maximum sequence length into every training example. Without sequence-aware partial projection at the storage layer, simpler models are forced to consume and transport the entire monolithic UIH data, leading to read amplification and making it prohibitive to scale UIH for flagship models without unfairly impacting the rest of the fleet.

\section{Methodology}
\label{sec:methodology}

To circumvent this storage and I/O wall, we pivot from the storage-heavy pre-materialization ``Fat Row'' paradigm to a compute-centric late materialization approach that stores UIH sequence features as lightweight version metadata referencing a normalized, versioned history, decoupling the storage footprint of UIH from the number of training examples. For this approach to be viable, it must ensure O2O consistency---reconstructing the exact inference-time UIH state during training---and preserve data freshness, as any regression in the availability of recent events would impair the model's ability to capture real-time user intent shifts. We first argue why late materialization is safe for UIH sequences (Section~\ref{sec:false_necessity}), then present the practical challenge of serving both streaming and batch training (Section~\ref{sec:streaming_batch}), and finally introduce the \emph{versioned late materialization} protocol (Section~\ref{sec:protocol}).

\subsection{The False Necessity of Pre-Materialization}
\label{sec:false_necessity}

The universality of the Fat Row architecture rests on an implicit thesis: to ensure O2O consistency and prevent future leakage, the complete UIH sequence must be physically snapshotted into every training example at inference time. This thesis conflates two independent claims: (C1)~that O2O consistency is a hard requirement, and (C2)~that physical pre-materialization is the \emph{necessary mechanism} to achieve it. Claim~C1 is unambiguously true. However, Claim~C2 is only sufficient, not necessary, and it is the sole driver of the infrastructure crisis described in Section~\ref{sec:motivation}.

Claim~C2 does not hold for UIH sequences due to a structural invariant: unlike general features whose values may be overwritten in place, UIH is an append-only, temporally ordered, immutable sequence---a user's history at time $t$ is deterministically defined as the set of events with $\texttt{timestamp} \leq t$, and individual events are never retroactively modified or deleted once written. This invariant has three consequences that collectively invalidate C2. First, \emph{reconstructibility}: the exact state of UIH at any historical inference time $t$ can be recovered from a single canonical copy via a temporal range scan, which requires no physical snapshot. Second, \emph{future leakage prevention by predicate}: the temporal predicate $\texttt{timestamp} \leq t$ is both necessary and sufficient to exclude post-inference events, providing the same guarantee as physical snapshotting without data duplication. Third, \emph{version metadata sufficiency}: logging a lightweight UIH sequence versioning feature ($O(1)$ per sample) replaces logging the entire sequence ($O(\text{seq\_length})$). In database terms, this admits an MVCC-style protocol~\cite{berenson2007critiqueansisqlisolation}---store canonical data once, log version metadata, and reconstruct state at read time---an application of late materialization~\cite{abadi2007materialization} to the recommendation training domain.

\subsection{Batch and Streaming Training}
\label{sec:streaming_batch}

Data freshness is a well-established driver of DLRM quality, enabling models to capture emerging trends and in-session user intent shifts. To satisfy freshness requirements while supporting large-scale experimentation, a unified training data generation pipeline is maintained for each of our products to serve both training paradigms. Online streaming training consumes a distributed real-time messaging stream~\cite{MetaScribe} to power latency-sensitive production models. The same stream is persisted in data warehouse tables~\cite{MetaHive} hourly to allow batch-based offline training for experimental iterations and new model warm-up.

Eliminating the ``Fat Row'' redundancy within this unified pipeline is non-trivial. A naive normalization approach, storing UIH features in a side table and joining them during an offline ETL process, fails for two reasons. First, it violates the unified ingestion model: online training workers cannot perform heavy analytical joins on live streams without breaching stringent latency SLAs. Second, it introduces an O2O consistency gap: if the offline join materializes a different UIH snapshot than the one observed during online ranking, the resulting training-serving skew degrades model quality. These constraints motivate the \emph{versioned late materialization} protocol presented in Section~\ref{sec:protocol}, which provides a unified UIH point-in-time reconstruction mechanism that is transparent to training job types.

\subsection{Versioned Late Materialization Protocol}
\label{sec:protocol}

\begin{figure*}[t]
  \centering
  \includegraphics[width=\linewidth]{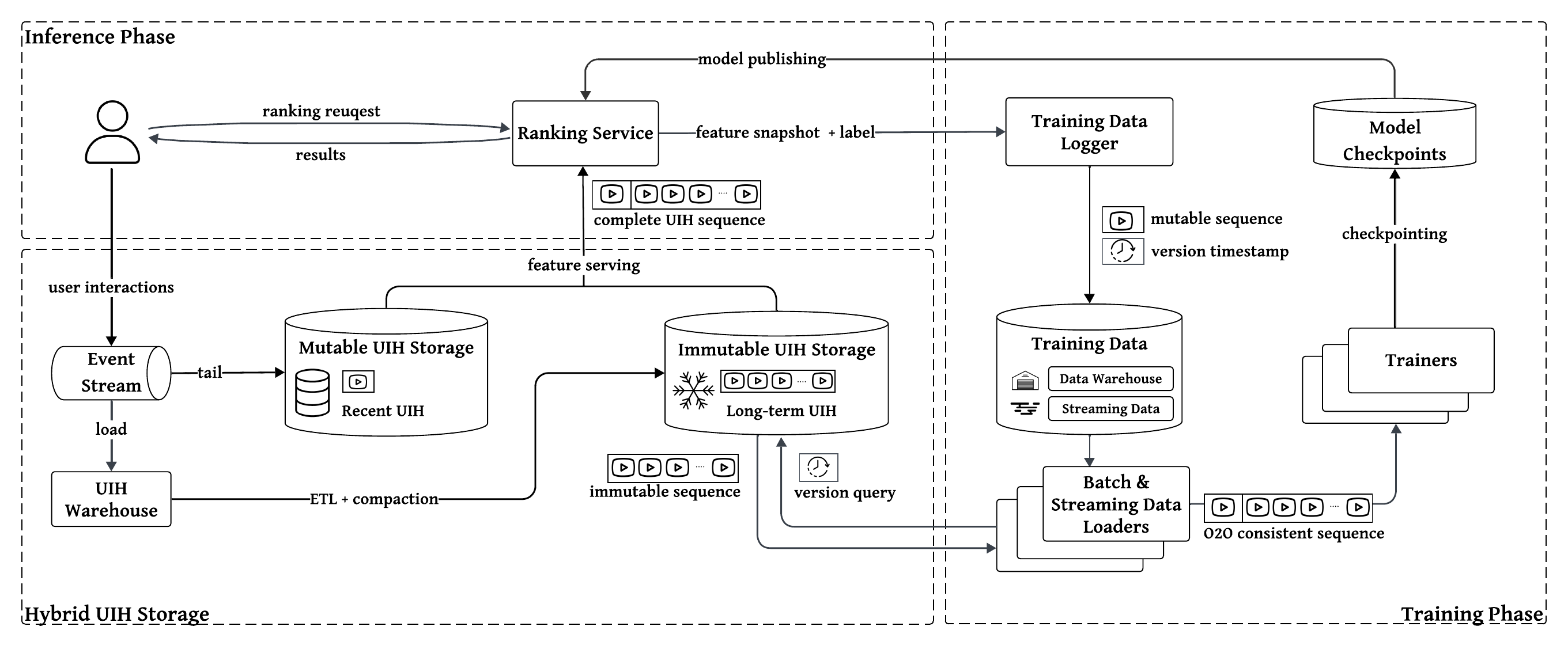}
  \caption{Versioned Late Materialization Protocol}
  \Description{Diagram illustrating the versioned late materialization protocol, showing inference-time snapshotting of mutable UIH and version metadata, immutable UIH storage organization, and training-time reconstruction via bounded range scan.}
  \label{fig:protocol}
\end{figure*}

As illustrated in Figure~\ref{fig:protocol}, the protocol is built upon the bifurcation of UIH into a Mutable UIH portion, which handles recent events comprising a small percentage of the UIH sequence, and an Immutable UIH portion for a user's long-term history (physical storage optimizations are detailed in Section~\ref{sec:productionization}). The protocol consists of lightweight inference-time snapshotting and versioned training-time late materialization.

\textbf{Inference-Time Snapshotting.} During the ranking phase, the system fetches sequences from both Mutable UIH and Immutable UIH to construct a complete UIH sequence for model inference. To minimize feature logging volume, instead of logging the snapshot of complete UIH at inference time, the system only persists the mutable UIH for each training example. For the massive immutable UIH portion, the system logs only the lightweight version metadata required for training-time UIH reconstruction: temporal boundaries (\texttt{start\_ts}, \texttt{end\_ts}), sequence length, and an optional checksum to verify the correctness of the reconstructed UIH during training. By logging compact versioned metadata rather than raw sequence data, we effectively decouple logging write-amplification from the total sequence length.

\textbf{Immutable Data Organization.} To facilitate efficient training-time retrieval, the immutable UIH storage organizes the history into subsequences, each keyed by its initial timestamp (\texttt{start\_ts}). This layout allows the training-time range scan query to precisely fetch the immutable UIH data within target temporal windows and sequence length, ensuring high-concurrency read performance even when 100+ training jobs access overlapping ranges.

\textbf{Training-Time Versioned Late Materialization.} During the training data preprocessing, the system performs a ``Time-Travel'' reconstruction to replicate the inference-time state:
\begin{enumerate}
  \item Extracts the version metadata and the recent sequence from the mutable UIH logged in the training example.
  \item Simultaneously applies the temporal boundaries from the logged version metadata to issue a bounded range scan against the immutable UIH storage.
  \item Concatenates the two components to reconstruct complete UIH features consistent with the inference-time snapshot.
\end{enumerate}

\textbf{Correctness Argument.} This protocol maintains O2O consistency by construction across both storage systems. The mutable UIH, comprising recent engagements immediately preceding the ranking request, is physically snapshotted and recorded at $T_{\text{request}}$, ensuring that no late-arrival events can contaminate this portion. For immutable UIH, the system issues a bounded range scan at training-time materialization using the temporal boundaries (\texttt{start\_ts}, \texttt{end\_ts}) logged at inference time; since the historical data stored in immutable UIH storage are static, the same bounded range scan yields identical results at both inference and training time. Any events loaded into immutable storage after $T_{\text{request}}$ (including interactions between $T_{\text{request}}$ and $T_{\text{train}}$) are excluded from the versioned window. In production, checksum-based validation between the Fat Row baseline and late-materialized sequences confirms reconstruction fidelity, empirically verifying the consistency guarantee. Importantly, since the reconstruction logic operates entirely within the data preprocessing layer and depends only on the logged version metadata, not on the training paradigm, the protocol can be seamlessly adopted by both streaming and batch training.

\section{Productionization}
\label{sec:productionization}

The \emph{versioned late materialization} protocol in Section~\ref{sec:methodology} eliminates the storage redundancy of the Fat Row paradigm and ensures Online-to-Offline consistency for training data correctness. However, productionizing this protocol within the production training ecosystem introduces two key systems challenges. First, hundreds of concurrent training jobs for a single product, spanning production models and experimental iterations across multiple training regions, all issue read queries against a shared UIH storage, creating a degree of read amplification that demands a highly scalable storage architecture (Section~\ref{sec:storage}). Second, shifting sequence construction to the training read-path requires each training data preprocessing worker to perform non-contiguous random reads from the immutable UIH storage, risking GPU starvation without aggressive optimization (Section~\ref{sec:materialization}).

\subsection{Scalable UIH Storage}
\label{sec:storage}

\subsubsection{Hybrid Storage Design}

The system bifurcates UIH into two storage systems to serve the divergent access patterns of real-time inference and high-concurrency training: a real-time mutable UIH store that captures the most recent engagements to provide second-level feature freshness, and an immutable UIH store that stores the vast majority of the UIH volume as long-term historical sequences populated via daily compaction jobs. The mutable UIH is accessed exclusively on the inference path; according to the versioned protocol, the mutable sequence is captured and logged into the training data at inference time, so training workers never query the mutable UIH. To support high-frequency updates without a Read-Modify-Write penalty, the mutable UIH store employs a fast merge operator that performs blind-write appends, deferring state resolution to read-time or background compaction, and a write-through cache co-located with the ranking service for minimal access latency. The retention window of the mutable store is coupled to the compaction cadence of the immutable store: events must remain in the mutable tier until the next offloaded compaction cycle (Section~\ref{sec:immutable}) consolidates them into the immutable tier, so a daily compaction schedule requires at least one day of mutable retention. The remainder of this section focuses on the immutable UIH, which dominates both the storage volume and the read traffic from the training fleet and the online ranking service.

\subsubsection{Read-Optimized Immutable Storage}
\label{sec:immutable}

The immutable UIH store serves as the normalized repository for long-term user interaction history, eliminating the systemic redundancy of the Fat Row paradigm. Its design is driven by a single objective: maximizing read throughput for range scans over versioned UIH sequences under massive concurrent training load while enabling multi-dimensional projection for heterogeneous model tenants.

\textbf{Offloaded compaction and optimal data layout.} The immutable UIH store has a predictable query pattern. Every query retrieves a contiguous subsequence of a user's history within a temporal range. This constrained access pattern admits a data layout that is provably optimal in I/O efficiency: one disk seek followed by purely sequential I/O per query. The immutable and read-only nature of historical UIH eliminates the need for write-optimized data layout \cite{patrick1996lsmt}, allowing a single-level storage layout that no general-purpose storage engine \cite{dong2021rocksdb} can match. To maintain this optimal layout, a daily ETL pipeline merges and compacts newly arrived UIH partitions with existing history into complete, chronologically ordered sequences, and generates pre-sorted storage files whose key ordering and sharding match the storage engine's topology. These pre-sorted files are then bulk-loaded into the immutable UIH store as a single-level layout, eliminating the multi-level compaction that LSM-based stores \cite{patrick1996lsmt} require to reorganize ingested data. This preserves the optimal data layout across daily refreshes, serving massive concurrent read traffic from the training fleet without compaction-induced write amplification or multi-level read amplification. Since each generation cycle rebuilds the full lookback window from source-of-truth data, it coalesces all temporal stripes for a given user into contiguous storage, ensuring that multi-stripe range scans remain purely sequential I/O rather than random reads scattered across multiple files.

\textbf{Multi-dimensional projection schema.} To support the heterogeneous requirements of different model tenants, the offline UIH compaction pipeline organizes each user's history by a multi-dimensional composite key with \texttt{user\_id}, \texttt{feature\_group}, and \texttt{subsequence\_timestamp}. This schema partitions a user's monolithic sequence into fixed-length temporal subsequences---analogous to horizontal stripes of user engagements---each stored as a separate row keyed by its start timestamp. This layout enables two dimensions of server-side projection pushdown that together eliminate the multi-tenant penalty described in Section~\ref{sec:union}. First, \emph{sequence length projection}: the system computes exactly how many stripes to scan for a model's target sequence length, so shorter-sequence tenants never over-fetch the full history. Second, \emph{feature group projection}: different models select only the feature groups relevant to their architecture, so simpler tenants never read the superset of all features. At the storage engine level, the system leverages an optimized multi-range scan with parallel I/O and prefetching to fetch multiple stripes in a single batched request, amortizing per-request latency and maximizing throughput for the co-located sequential reads enabled by the optimal single-layer data layout.

\textbf{Trait-aware columnar encoding.} The system employs a domain-specific columnar encoding format that organizes user history as a column-oriented matrix where rows represent chronologically ordered events and columns denote typed, heterogeneous traits (e.g., post ID, timestamp, event type, video watch time). Notably, these traits exhibit different density and value distributions. For example, post ID and timestamp are dense, while engagement signals like comment and share are highly sparse. Within each horizontal stripe of the user sequence, the system exploits this heterogeneity through a trait-aware columnar encoding~\cite{liao2024bullioncolumnstoremachine}---delta-encoding for timestamps, compact presence bitmaps for likes, or dictionary compression for categorical metadata. Besides storage and I/O efficiency benefits, the columnar layout enables selective decoding: at materialization time, it decodes only the subset of traits required by the model's feature specification, skipping irrelevant columns entirely at the byte level. This additional secondary-level projection complements the \emph{sequence-length} and \emph{feature-group} projections, minimizing I/O bandwidth and decoding overhead.

\subsection{Training-Time Materialization}
\label{sec:materialization}

\subsubsection{Disaggregated Data Preprocessing}
\label{sec:dpp}

Late materialization shifts the burden of ultra-long sequence reconstruction---I/O-intensive joins and decoding against immutable UIH sequences---from pre-materialization pipelines into the training read-path, risking GPU starvation where accelerators idle waiting for training data to be loaded. To address this, we adopt a disaggregated Data PreProcessing (DPP) architecture~\cite{zhao2022understanding} that decouples training data loading and preprocessing from the training loop by offloading materialization to independently scalable preprocessing clusters. This disaggregated architecture enables elastic scaling: the system monitors the job-level GPU starvation percentage (accelerator idle time) and worker waste percentage (CPU idle time) in real time, and automatically provisions supplemental DPP workers for jobs with complex UIH reconstruction, ensuring that training throughput remains compute-bound by the GPUs rather than bottlenecked by the non-contiguous I/O latency and decoding of sequence materialization. Moreover, modern GPUs require large batch sizes for peak utilization, but materializing ultra-long sequences renders DPP worker threads heavily memory-bound. This disaggregated architecture achieves large training batches within each DPP worker's memory boundary by trainer-side rebatching. DPP workers process smaller base batches that fit within physical memory constraints, which are then asynchronously buffered, merged, and reshuffled by the trainer-side DPP client to satisfy the model's full batch requirements. This reduction in preprocessing granularity enables higher thread concurrency and CPU utilization per worker. In practice, tuning the DPP base batch size to balance memory pressure against thread-level parallelism yields a 15\% improvement in per-worker data preprocessing throughput.

\subsubsection{Masking Latency}
\label{sec:latencymasking}

To mask the non-contiguous I/O latency of immutable sequence lookups during training, we integrate an efficient vectorized execution engine~\cite{pedro2022velox} as the query engine on DPP workers. The core operator is a specialized index join where the \emph{probe-side} is the primary training table, and the \emph{build-side} is the immutable UIH store. Unlike conventional joins in OLAP workloads, this operator is designed to hide remote storage latency:

\textbf{Pipelined I/O Prefetching.} For each probe-side training batch, the operator extracts join keys and issues a bulk multi-range scan request against the immutable sequence storage. Without waiting for the response, it immediately prefetches the next probe-side batch, overlapping the immutable storage read for batch $N$ with the primary training table read for batch $N+1$. Since these latencies are comparable, concurrent execution effectively masks the sequence lookup latency without stalling the pipeline, achieving an additional 10\% improvement in per-worker data preprocessing throughput on models with heavy UIH sequence lookup.

\textbf{Partial Projection Pushdown.} To mitigate I/O amplification across a heterogeneous fleet of model tenants, the system exposes each model's UIH sequence length and feature group requirements to the query engine, which pushes these projections down to the immutable UIH storage, enabling the storage optimizations described in Section~\ref{sec:immutable} to execute a partial range scan that retrieves only the relevant feature subsets and the specific temporal window requested by the model---avoiding the multi-tenant over-fetch penalty described in Section~\ref{sec:union}.

\subsubsection{Data-Affinity I/O Optimization}
\label{sec:affinity}

Streaming training jobs naturally achieve a high block cache hit rate on the immutable UIH store, as concurrent trainers process the same recent temporal partitions. However, offline batch training jobs---including experimental iterations and warm-up runs---access disparate historical partitions with minimal cache locality, requiring explicit I/O optimization.

The system employs two complementary data-affinity strategies for offline batch training. First, during ingestion from real-time streams into the data warehouse for batch training, the system clusters training examples into buckets keyed by users. This groups a user's training examples from the same hourly window into a single batch, allowing DPP workers to perform the UIH reconstruction for all temporally-adjacent examples of the same user with a single immutable sequence lookup, effectively amortizing retrieval workload across session-level interactions. Second, the system enforces symmetric sharding between the primary training data and the immutable UIH storage using an identical hash partitioning scheme with a shared partition key, ensuring that all UIH sequence lookups within a data loading batch map to the same storage shard and eliminating the network fanout that would otherwise bottleneck high-concurrency read-paths. Together, these two optimizations reduce the overall read bandwidth from batch training jobs by 60\% and increase the per-worker data processing throughput by 28\%.

\subsection{Enabling Model Scaling}

Removing the data wall via late materialization is necessary but not sufficient---training and serving ultra-long sequences also demands model-side compute efficiency. ULTRA-HSTU~\cite{ding26ultra} and VISTA~\cite{chen2026massivememorizationhundredstrillions} provide the complementary efficiency stack through a two-stage framework, semi-local attention, load-balanced stochastic sampling, and kernel-level optimizations that together enable higher throughput at doubled sequence length than the original baseline. On the data side, the immutable storage supports a \emph{layered sequence layout} that partitions UIH into density-aware feature groups---e.g., dense view events with short lookback versus sparse explicit actions with long lookback---powered by the multi-dimensional projection pushdown in Section~\ref{sec:immutable} to serve each group independently for maximizing signal value within the same sequence length. Data infrastructure and model efficiency are complementary investments that amplify each other when combined~\cite{ding26ultra, li2025fm}.

Beyond compute efficiency, the offloaded compaction design of the immutable UIH storage (Section~\ref{sec:immutable}) provides important operational benefits that accelerate the model experiment lifecycle. Since each daily compaction regenerates the entire lookback window, it naturally enforces \emph{right-to-delete} compliance: engagements associated with deleted content or accounts are idempotently scrubbed during every bulk load cycle, eliminating the need for expensive retroactive patching of historical sequences. Equally important, when researchers introduce new SideInfo features (e.g., content category, creator attributes) or deprecate unused ones, a single pipeline run regenerates the complete lookback window with the updated schema---enabling rapid offline experimentation without the multi-day backfill latency that incremental append-only systems would require. This agility in feature iteration is essential for improving model quality, as the optimal set of UIH features co-evolves with model architecture and sequence length.

\section{Evaluation}
\label{sec:evaluation}

\subsection{System Efficiency}
\label{sec:eval_efficiency}

\begin{table*}[t]
  \caption{System Efficiency of Versioned Late Materialization (Baseline = Fat Row Paradigm)}
  \label{tab:resource-efficiency}
  \begin{tabular}{lcccccccc}
    \toprule
    \multirow{2}{*}{Model Tenant} & \multirow{2}{*}{\begin{tabular}[c]{@{}c@{}}UIH Seq.\\ Length\end{tabular}} & \multirow{2}{*}{\begin{tabular}[c]{@{}c@{}}Number of\\ Training Jobs\end{tabular}} &
    \multirow{2}{*}{\begin{tabular}[c]{@{}c@{}}Primary Write\\ Bandwidth$^1$\end{tabular}} &
    \multirow{2}{*}{\begin{tabular}[c]{@{}c@{}}Primary Read\\ Bandwidth$^2$\end{tabular}} &
    \multicolumn{2}{c}{Seq.\ Lookup Bandwidth$^3$} & \multirow{2}{*}{\begin{tabular}[c]{@{}c@{}}Per-Batch Data\\ Loading Latency\end{tabular}} \\
    \cmidrule(lr){6-7}
    & & & & & Streaming & Batch & \\
    \midrule
    Model A & Long & Low & \multirow{3}{*}{$-$46.2\%} & $-$70.3\% & +62.7\% & +24.6\% & $+$9.7\% \\
    Model B & Mid & High &  & $-$50.9\% & +16.2\% & +6.5\% & -26.4\% \\
    Model C & Short & High &  & $-$47.7\% & +8.7\% & +3.4\% & -36.2\% \\
    \bottomrule
  \end{tabular}
  \vspace{1mm}
  \begin{flushleft}
    \small $^1$ Write bandwidth reduction is shared across 3 model tenants using the shared training dataset.\\
    \small $^2$ Primary training data includes all labels and features, excluding the normalized immutable UIH sequences.\\
    \small $^3$ UIH sequence lookup bandwidth from the immutable UIH store, reported as a percentage relative to the baseline primary read bandwidth.\\
  \end{flushleft}
\end{table*}

Table~\ref{tab:resource-efficiency} summarizes the system metrics of \emph{versioned late materialization} evaluated on three model tenants that share a union training dataset of a recommendation platform. All metrics are measured under streaming training unless explicitly marked. By storing UIH sequences once in a normalized tier rather than duplicating them into every training example, the system eliminates the $K$-fold write amplification inherent in denormalized Fat Rows, yielding a 46.2\% reduction in primary write bandwidth of the shared training dataset. On the read side, removing the UIH payload from the primary training data reduces per-job read bandwidth by 47--70\%, with Models~B (mid sequence) and~C (short sequence) further benefiting from projection pushdown (Section~\ref{sec:materialization}): they fetch only the required sequence lengths rather than the full long sequence, eliminating the multi-tenant read amplification penalty (Section~\ref{sec:union}).

Late materialization introduces sequence lookup as new I/O against the immutable UIH store. The streaming training of Model~A (long sequence) has the highest sequence lookup bandwidth at 62.7\% relative to the baseline primary read, narrowed from the 70.3\% primary read reduction by the trait-aware columnar encoding (Section~\ref{sec:immutable}) that achieves a higher encoding density. Although 62.7\% appears to nearly offset the 70.3\% savings in raw bandwidth, the single-level, compaction-free immutable storage delivers 3.4$\times$ higher read throughput per unit of host resource than the append-only primary training data storage, so the effective host resource footprint of the 62.7\% lookup bandwidth is substantially lower than the resources freed by the 70.3\% primary read reduction. For batch training, the data-affinity optimization (Section~\ref{sec:affinity}) further amortizes sequence lookups across temporally adjacent examples of the same user, reducing lookup bandwidth by approximately 60\%. Average per-batch data loading latency governs per-worker preprocessing throughput and thus the number of DPP workers each training job provisions to keep GPUs saturated. Model~A sees a modest 9.7\% increase from training-time sequence reconstruction, which is minimized by pipelined I/O prefetching (Section~\ref{sec:latencymasking}) and absorbed by elastic DPP scaling (Section~\ref{sec:dpp}) to avoid GPU starvation. Models~B and~C achieve 26--36\% latency improvements through projection pushdown. The system-level efficiency is further amplified by the job distribution: Model~A's higher per-job GPU and memory footprint results in longer iteration cycles and lower job concurrency, while Models~B and~C account for the majority of concurrent training jobs and realize the largest per-job improvements in both read bandwidth and data loading latency.

\subsection{Model Performance}
\label{sec:eval_model}

\begin{figure}[t]
  \centering
  \includegraphics[width=\linewidth]{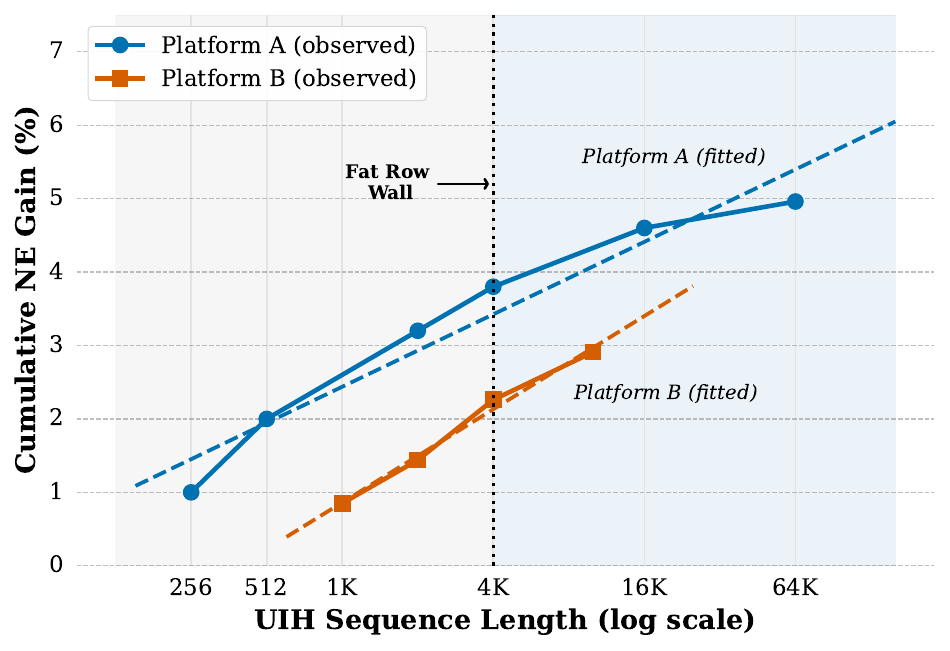}
  \caption{NE improvement of UIH sequence length scaling on two production recommendation surfaces.}
  \Description{Line chart showing NE improvement vs sequence length for two platforms with the Fat Row Wall marked.}
  \label{fig:ne_scaling}
\end{figure}

We evaluated the impact of sequence length scaling on the late-stage ranking models of two production recommendation platforms using Normalized Entropy (NE)~\cite{he2014practical}. As shown in Figure~\ref{fig:ne_scaling}, both platforms exhibit a monotonically improving NE as the UIH sequence length increases under the \emph{versioned late materialization} infrastructure. Across the overlapping range where both paradigms are deployable (256--4K), late materialization achieves NE on par with the Fat Row baseline, confirming that training-time sequence reconstruction does not introduce model quality degradation. We define the \emph{Fat Row Wall} as the sequence length at which the ratio of data supporting service resource usage to GPU training power exceeds 0.75. In our production environment, this wall is reached at approximately 4K (Figure~\ref{fig:ne_scaling}), beyond which the $K$-fold amplification (Section~\ref{sec:motivation}) makes further scaling under the Fat Row paradigm difficult. The \emph{versioned late materialization} architecture breaks through this wall: Platform A gains an additional 1.2\% cumulative NE improvement from 4K to 64K (total >5\%), and Platform B gains 0.65\% from 4K to 10K. At this production scale, where a 1\% NE improvement is considered a success, these gains, enabled by \emph{versioned late materialization}, would be impractical to achieve under the Fat Row paradigm within the same infrastructure envelope.

\begin{table}[t]
    \caption{Online A/B testing results of sequence length scaling enabled by versioned late materialization on two production recommendation platforms.}
\label{tab:online-results}
\centering
\begin{tabular}{lccc}
  \toprule
  \textbf{Product} & \textbf{Seq.\ Length} & \textbf{Metric} & \textbf{Relative Gain} \\
  \midrule
  \multirow{4}{*}{\textbf{Platform A}} & \multirow{4}{*}{4K $\rightarrow$ 16K}
    & Metric-Topline   & +0.22\% \\
    & & Metrics-C        & +4.1\%  \\
    & & Metrics-E-1        & +2.3\%  \\
    & & Metrics-E-2        & +4.3\%  \\
  \midrule
  \multirow{4}{*}{\textbf{Platform B}} & \multirow{4}{*}{4K $\rightarrow$ 10K}
    & Metric-Topline   & +0.14\%  \\
    & & Metrics-C        & +0.79\%  \\
    & & Metrics-E-3        & +1.4\%  \\
    & & Metrics-E-4        & +1.7\%  \\
  \bottomrule
\end{tabular}
\end{table}
The impact of sequence length scaling aided by \emph{versioned late materialization} is further validated through rigorous online A/B testing on these two recommendation platforms, as summarized in Table~\ref{tab:online-results}. Following prior work~\cite{ding26ultra}, we report anonymized metrics across three categories: Metric-Topline (the platform's north-star engagement metric), Metrics-C (consumption-oriented metrics such as time spent), and Metrics-E (explicit engagement signals such as reshares, comments, and likes). For each platform, the target sequence length---4K\,$\rightarrow$\,16K for Platform A and 4K\,$\rightarrow$\,10K for Platform B---is selected to balance model quality gains against the marginal GPU and data supporting service usage at each scaling step. Notably, these gains are demonstrated across platforms powered by distinct model architectures, confirming that the sequence infrastructure's benefits are broadly applicable.

\section{Related Work}
\label{sec:related}

As discussed in Section~\ref{sec:intro}, a rich body of work has driven UIH scaling to $10^5$+ events through increasingly sophisticated model architectures~\cite{zhou2018deep, pi2020search, chen2021end, zhai2024actions, si2024twin, li2025fm, ding26ultra}. However, these efforts focus exclusively on model architecture and attention efficiency, largely treating the underlying data availability as a given. Our work addresses the complementary and often bottlenecking challenge: the data infrastructure that determines how far the sequence length can be scaled efficiently for training.

General-purpose ML data pipeline frameworks such as tf.data~\cite{murray2021tfdata} and disaggregated preprocessing architectures~\cite{zhao2022understanding, um2023fastflow} address training data loading and preprocessing bottlenecks, while feature stores~\cite{polyzotis2017data} provide unified offline/online feature serving to mitigate training-serving skew. These systems target broad ML workloads but do not address the domain-specific challenge of DLRM training, where UIH sequences dominate storage and I/O footprint with $K$-fold redundancy across multi-tenant fleets. More recently, ROO~\cite{guo2025request} introduces request-only optimization for session-level training data generation in recommendation systems, reducing redundancy in non-sequential features across impressions within a session. However, UIH sequence redundancy---the dominant system resource driver identified in Section~\ref{sec:motivation}---remains untouched, as ROO does not normalize or deduplicate the long-history sequences that are physically duplicated in every training example.

Late materialization---deferring reconstruction until query results are needed---is a well-established technique in columnar database query processing~\cite{abadi2007materialization} that reduces I/O by avoiding premature assembly of wide rows. Multi-version concurrency control (MVCC)~\cite{bernstein1983mvcc} is similarly foundational in transactional databases, enabling concurrent readers to observe consistent snapshots without blocking writers. While both techniques are mature in the database community, none of these systems address the domain-specific challenges of recommendation training: the interplay of streaming and batch training modes with distinct temporal semantics, the need for future-leakage prevention across both modes, and multi-tenant sequence projection over variable-length histories. Our work adapts these design principles to the recommendation systems ML training domain---storing normalized sequences once and reconstructing them just-in-time during training via lightweight version metadata, while introducing a bifurcated consistency protocol that addresses these domain-specific requirements.

\section{Conclusion}
\label{sec:conclusion}
We presented \emph{versioned late materialization}, a data infrastructure paradigm that eliminates the $K$-fold storage and I/O redundancy of the Fat Row approach by exploiting a structural invariant of User Interaction History: as an append-only, temporally ordered sequence, its state at any point in time can be reconstructed from a single canonical copy via lightweight version metadata rather than physical duplication. Deployed across multiple production recommendation surfaces, the system not only reduces data infrastructure resource usage but, more importantly, enables UIH sequence scaling to lengths impractical under the Fat Row paradigm, unlocking significant model quality gains. These results demonstrate that data infrastructure is a first-class scaling lever for recommendation quality, complementary to advances in model architecture.

\begin{acks}
This work would not be possible without the work from the following contributors (alphabetical order): Prashasti Baid, Renqin Cai, Xianjie Chen, Huihui Cheng, Patrick Cullen, Chaitanya Datye, Delia David, Fei Ding, Qi Ding, Qin Ding, Omkar Gawde, Sathya Gunasekar, Xinyao Hu, Yanzun Huang, Parichay Kapoor, Manos Karpathiotakis, Hung-Ching Lee, Hongwei Li, Jiahao Liang, Xueyao Liang, Linbin Luo, Yun Mao, Franco Mo, Luning Pan, Pedro Eugenio Rocha Pedreira, Harsha Rastogi, Kai Ren, Aleksandr Shatrovskii, Hongzheng Shi, Yu Shi, Ruohan Sun, Mike Trepanier, Igor Vodov, Shanyue Wan, Hao Wang, Junjie Wang, Zehui Wang, Xiang Xiao, Kostas Xirogiannopoulos, Shuo Xu, Hon Yan, Tao Yang, Stanley Yao, Ximing Yu, Bill Zhang, Xin Zhang, Zhang Zhang, Zhenyuan Zhao, Charles Zheng, Josh Zhou, Yupeng Zou.
\end{acks}

\bibliographystyle{ACM-Reference-Format}
\bibliography{seq-scale}

@inproceedings{zhou2018deep,
  title={Deep Interest Network for Click-Through Rate Prediction},
  author={Zhou, Guorui and Mou, Na and Fan, Ying and Pi, Qi and Bian, Weijie and Zhou, Changhua and Zhu, Xiaoqiang and Gai, Kun},
  booktitle={Proceedings of the 24th ACM SIGKDD International Conference on Knowledge Discovery \& Data Mining},
  pages={1059--1068},
  year={2018}
}

@article{pi2020search,
  title={Search-based User Interest Modeling with Lifelong Sequential Behavior Data for Click-Through Rate Prediction},
  author={Pi, Qi and Zhou, Guorui and Zhang, Yujing and Wang, Zhe and Ren, Lejian and Fan, Ying and Zhu, Xiaoqiang and Gai, Kun},
  journal={arXiv preprint arXiv:2006.05639},
  year={2020}
}

@article{chen2021end,
  title={End-to-End User Behavior Retrieval in Click-Through Rate Prediction Model},
  author={Chen, Qiwei and Pei, Changhua and Lv, Shuguang and Li, Cheng and Ge, Jian and Ou, Wenwu},
  journal={arXiv preprint arXiv:2108.04468},
  year={2021}
}

@inproceedings{zhai2024actions,
  title={Actions Speak Louder than Words: Trillion-Parameter Sequential Transducers for Generative Recommendations},
  author={Zhai, Jiaqi and Liao, Lucy and Liu, Xing and Wang, Yueming and Li, Rui and Cao, Xuan and Gao, Leon and Gong, Zhaojie and Gu, Fangda and He, Jiayuan and Lu, Yinghai and Shi, Yu},
  booktitle={Proceedings of the 41st International Conference on Machine Learning},
  pages={58484--58509},
  year={2024},
  organization={PMLR}
}

@inproceedings{si2024twin,
  title={Twin v2: Scaling ultra-long user behavior sequence modeling for enhanced ctr prediction at kuaishou},
  author={Si, Zihua and Guan, Lin and Sun, ZhongXiang and Zang, Xiaoxue and Lu, Jing and Hui, Yiqun and Cao, Xingchao and Yang, Zeyu and Zheng, Yichen and Leng, Dewei and others},
  booktitle={Proceedings of the 33rd ACM International Conference on Information and Knowledge Management},
  pages={4890--4897},
  year={2024}
}

@misc{ding26ultra,
  title={Bending the Scaling Law Curve in Large-Scale Recommendation Systems},
  author={Qin Ding and Kevin Course and Linjian Ma and Jianhui Sun and Ruochen Liu and Zhao Zhu and Chunxing Yin and Wei Li and Dai Li and Yu Shi and Xuan Cao and Ze Yang and Han Li and Xing Liu and Bi Xue and Hongwei Li and Rui Jian and Daisy Shi He and Jing Qian and Matt Ma and Qunshu Zhang and Rui Li},
  year={2026},
  eprint={2602.16986},
  archivePrefix={arXiv},
  primaryClass={cs.IR},
  url={https://arxiv.org/abs/2602.16986},
}

@misc{li2025fm,
  title={Realizing Scaling Laws in Recommender Systems: A Foundation-Expert Paradigm for Hyperscale Model Deployment},
  author={Dai Li and Kevin Course and Wei Li and Hongwei Li and Jie Hua and Yiqi Chen and Zhao Zhu and Rui Jian and Xuan Cao and Bi Xue and Yu Shi and Jing Qian and Kai Ren and Matt Ma and Qunshu Zhang and Rui Li},
  year={2025},
  eprint={2508.02929},
  archivePrefix={arXiv},
  primaryClass={cs.IR},
  url={https://arxiv.org/abs/2508.02929},
}

@inproceedings{lyu2025dv365,
  title={DV365: Extremely Long User History Modeling at Instagram},
  author={Lyu, Wenhan and Tyagi, Devashish and Yang, Yihang and Li, Ziwei and Somani, Ajay and Shanmugasundaram, Karthikeyan and Andrejevic, Nikola and Adeputra, Ferdi and Zeng, Curtis and Singh, Arun K and others},
  booktitle={Proceedings of the 31st ACM SIGKDD Conference on Knowledge Discovery and Data Mining V. 2},
  pages={4717--4727},
  year={2025}
}

@inproceedings{chai2025longer,
  title={Longer: Scaling up long sequence modeling in industrial recommenders},
  author={Chai, Zheng and Ren, Qin and Xiao, Xijun and Yang, Huizhi and Han, Bo and Zhang, Sijun and Chen, Di and Lu, Hui and Zhao, Wenlin and Yu, Lele and others},
  booktitle={Proceedings of the Nineteenth ACM Conference on Recommender Systems},
  pages={247--256},
  year={2025}
}

@inproceedings{liu2025longterm,
  title={User Long-Term Multi-Interest Retrieval Model for Recommendation},
  author={Meng, Yue and Guo, Cheng and Hu, Xiaohui and Deng, Honghu and Cao, Yi and Liu, Tong and Zheng, Bo},
  booktitle={Proceedings of the Nineteenth ACM Conference on Recommender Systems},
  pages={1112--1116},
  year={2025}
}

@article{ren2025longretriever,
  title={LongRetriever: Towards Ultra-Long Sequence based Candidate Retrieval for Recommendation},
  author={Ren, Qin and Chai, Zheng and Xiao, Xijun and Zheng, Yuchao and Wu, Di},
  journal={arXiv preprint arXiv:2508.15486},
  year={2025}
}

@article{guo2025request,
  title={Request-Only Optimization for Recommendation Systems},
  author={Guo, Liang and Li, Wei and Liao, Lucy and Cheng, Huihui and Zhang, Rui and Shi, Yu and Wang, Yueming and Huang, Yanzun and Zhai, Keke and Wang, Pengchao and others},
  journal={arXiv preprint arXiv:2508.05640},
  year={2025}
}

@article{murray2021tfdata,
  title={tf.data: A Machine Learning Data Processing Framework},
  author={Murray, Derek G and Simsa, Jiri and Klimovic, Ana and Indyk, Ihor},
  journal={Proceedings of the VLDB Endowment},
  volume={14},
  number={12},
  year={2021}
}

@inproceedings{zhao2022understanding,
  title={Understanding Data Storage and Ingestion for Large-Scale Deep Recommendation Model Training},
  author={Zhao, Mark and Tirmazi, Niket and Erber, Jiyan and Ihm, Skye and Minnich, Aarti and Nair, Shashank and Sun, Dawn},
  booktitle={Proceedings of the 49th Annual International Symposium on Computer Architecture},
  pages={1042--1057},
  year={2022}
}

@article{um2023fastflow,
  title={FastFlow: Accelerating Deep Learning Model Training with Smart Offloading of Input Data Pipeline},
  author={Um, Taegeon and others},
  journal={Proceedings of the VLDB Endowment},
  volume={16},
  number={5},
  year={2023}
}

@inproceedings{polyzotis2017data,
  title={Data Management Challenges in Production Machine Learning},
  author={Polyzotis, Neoklis and Roy, Sudip and Whang, Steven Euijong and Zinkevich, Martin},
  booktitle={Proceedings of the 2017 ACM International Conference on Management of Data},
  year={2017}
}

@inproceedings{abadi2007materialization,
  title={Materialization Strategies in a Column-Oriented DBMS},
  author={Abadi, Daniel J and Myers, Daniel S and DeWitt, David J and Madden, Samuel},
  booktitle={Proceedings of the 23rd IEEE International Conference on Data Engineering},
  year={2007}
}

@inproceedings{he2014practical,
  title={Practical Lessons from Predicting Clicks on Ads at Facebook},
  author={He, Xinran and Pan, Junfeng and Jin, Ou and Xu, Tianbing and Liu, Bo and Xu, Tao and Shi, Yanxin and Atber, Antoine and Herbrich, Ralf and Bowers, Stuart and others},
  booktitle={Proceedings of the Eighth International Workshop on Data Mining for Online Advertising},
  pages={1--9},
  year={2014}
}

@article{pedro2022velox,
  title={Velox: Meta's Unified Execution Engine},
  author={Pedreira, Pedro and Erber, Orri and Kandula, Masha and Haas, Kevin and Hao, Yolanda and Gruenheid, Anja and Nair, Deepak and Liu, Hao and Zhu, Huameng and Fan, Wenlei and others},
  journal={Proceedings of the VLDB Endowment},
  volume={15},
  number={12},
  pages={3372--3384},
  year={2022}
}

@article{patrick1996lsmt,
author = {O’Neil, Patrick and Cheng, Edward and Gawlick, Dieter and O’Neil, Elizabeth},
title = {The log-structured merge-tree (LSM-tree)},
year = {1996},
issue_date = {Jun 1996},
publisher = {Springer-Verlag},
address = {Berlin, Heidelberg},
volume = {33},
number = {4},
issn = {0001-5903},
url = {https://doi.org/10.1007/s002360050048},
doi = {10.1007/s002360050048},
journal = {Acta Inf.},
month = jun,
pages = {351–385},
numpages = {35}
}

@article{MetaHive,
  author = {Thusoo, Ashish and Sarma, Joydeep and Jain, Namit and Shao, Zheng and Chakka, Prasad and Anthony, Suresh and Liu, Hao and Wyckoff, Pete and Murthy, Raghotham},
  year = {2009},
  month = {08},
  pages = {1626-1629},
  title = {Hive - A Warehousing Solution Over a Map-Reduce Framework.},
  volume = {2},
  journal = {PVLDB},
  doi = {10.14778/1687553.1687609}
}

@article{MetaScribe,
  author = {Karpathiotakis, Manos and Rizopoulos, Vlassios and Kahveci, Basri and Carotti, Tiziano and Gelum, Artem and Nada, Hazem and Dolgov, Yuri},
  year = {2025},
  month = {09},
  pages = {4817-4830},
  title = {Scribe: How Meta Transports Terabytes per Second in Real Time},
  volume = {18},
  journal = {Proceedings of the VLDB Endowment},
  doi = {10.14778/3750601.3750607}
}

@inproceedings{sculley2015hidden,
  author = {Sculley, D. and Holt, Gary and Golovin, Daniel and Davydov, Eugene and Phillips, Todd and Ebner, Dietmar and Chaudhary, Vinay and Young, Michael and Crespo, Jean-Francois and Dennison, Dan},
  title = {Hidden technical debt in Machine learning systems},
  year = {2015},
  publisher = {MIT Press},
  address = {Cambridge, MA, USA},
  booktitle = {Proceedings of the 29th International Conference on Neural Information Processing Systems - Volume 2},
  pages = {2503–2511},
  numpages = {9},
  location = {Montreal, Canada},
  series = {NIPS'15}
}

@article{bernstein1983mvcc,
  author = {Bernstein, Philip and Goodman, Nathan},
  year = {1983},
  month = {12},
  pages = {465-483},
  title = {Multiversion Concurrency Control - Theory and Algorithms.},
  volume = {8},
  journal = {ACM Trans. Database Syst.},
  doi = {10.1145/319996.319998}
}

@misc{chen2026massivememorizationhundredstrillions,
  title={Massive Memorization with Hundreds of Trillions of Parameters for Sequential Transducer Generative Recommenders},
  author={Zhimin Chen and Chenyu Zhao and Ka Chun Mo and Yunjiang Jiang and Jane H. Lee and Khushhall Chandra Mahajan and Ning Jiang and Kai Ren and Jinhui Li and Wen-Yun Yang},
  year={2026},
  eprint={2510.22049},
  archivePrefix={arXiv},
  primaryClass={cs.IR},
  url={https://arxiv.org/abs/2510.22049},
}

@article{dong2021rocksdb,
  author = {Dong, Siying and Kryczka, Andrew and Jin, Yanqin and Stumm, Michael},
  title = {RocksDB: Evolution of Development Priorities in a Key-value Store Serving Large-scale Applications},
  year = {2021},
  issue_date = {November 2021},
  publisher = {Association for Computing Machinery},
  address = {New York, NY, USA},
  volume = {17},
  number = {4},
  issn = {1553-3077},
  url = {https://doi.org/10.1145/3483840},
  doi = {10.1145/3483840},
  journal = {ACM Trans. Storage},
  month = oct,
  articleno = {26},
  numpages = {32},
}

@misc{liao2024bullioncolumnstoremachine,
  title={Bullion: A Column Store for Machine Learning},
  author={Gang Liao and Ye Liu and Jianjun Chen and Daniel J. Abadi},
  year={2024},
  eprint={2404.08901},
  archivePrefix={arXiv},
  primaryClass={cs.DB},
  url={https://arxiv.org/abs/2404.08901},
}

@misc{berenson2007critiqueansisqlisolation,
  title={A Critique of ANSI SQL Isolation Levels}, 
  author={Hal Berenson and Phil Bernstein and Jim Gray and Jim Melton and Elizabeth O'Neil and Patrick O'Neil},
  year={2007},
  eprint={cs/0701157},
  archivePrefix={arXiv},
  primaryClass={cs.DB},
  url={https://arxiv.org/abs/cs/0701157}, 
}

\end{document}